\documentclass[12pt]{article}

\usepackage{fancybox}

\usepackage{cite}
\usepackage{float}
\usepackage{amsfonts}
\usepackage{amsmath}
\usepackage{amsbsy}
\usepackage{graphicx}
\usepackage{amssymb}
\usepackage{amsthm}
\usepackage{bm}
\usepackage{epsfig}
\usepackage{latexsym}
\usepackage{pdflscape}
\usepackage{color}
\usepackage{here}
\usepackage{graphicx}
\numberwithin{equation}{section}

\def\llp{(\!(}
\def\rrp{)\!)}
\def\llb{[[}
\def\rrb{]]}

\begin{document}

\renewcommand{\thefootnote}{\fnsymbol{footnote}}

\begin{titlepage}
\begin{flushright}
{\footnotesize OCU-PHYS 442, YITP-16-18}
\end{flushright}
\bigskip
\begin{center}
{\LARGE\bf Orientifold ABJM Matrix Model:\\[8pt]
Chiral Projections and Worldsheet Instantons}\\
\bigskip\bigskip
{\large 
Sanefumi Moriyama\footnote{\tt moriyama@sci.osaka-cu.ac.jp}
\quad and \quad
Tomoki Nosaka\footnote{\tt nosaka@yukawa.kyoto-u.ac.jp}
}\\
\bigskip
${}^*${\it Department of Physics, Graduate School of Science,
Osaka City University\\
3-3-138 Sugimoto, Sumiyoshi, Osaka 558-8585, Japan}\\
\medskip
${}^{*\dagger}${\it Yukawa Institute for Theoretical Physics,
Kyoto University\\
Kitashirakawa-Oiwakecho, Sakyo, Kyoto 606-8502, Japan}
\end{center}

\bigskip

\begin{abstract}
We study the partition function of the orientifold ABJM theory, which is a superconformal Chern-Simons theory associated with the orthosymplectic supergroup.
We find that the partition function associated with any orthosymplectic supergroup can be realized as the partition function of a Fermi gas system whose density matrix is identical to that associated with the corresponding unitary supergroup with a projection to the even or odd chirality.
Furthermore we propose an identity which gives directly all of the Gopakumar-Vafa invariants for the worldsheet instanton effects in the chirally projected theories.

\bigskip\bigskip

\centering\includegraphics[scale=0.6]{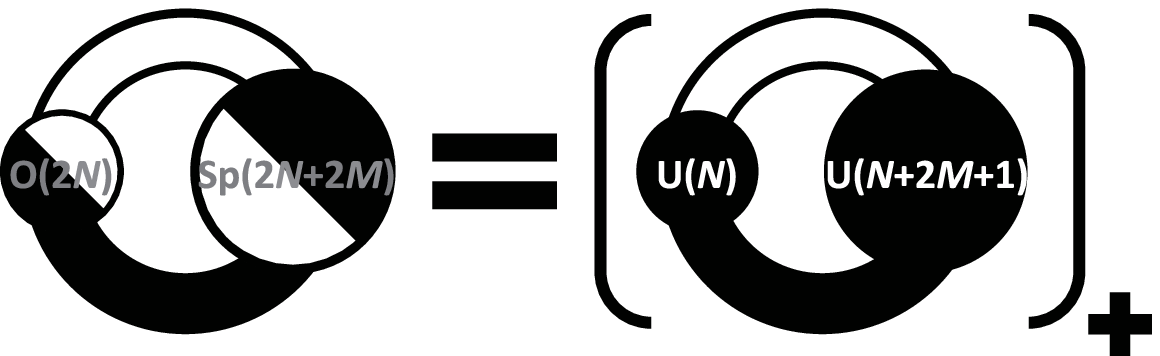}\quad\includegraphics[scale=0.6]{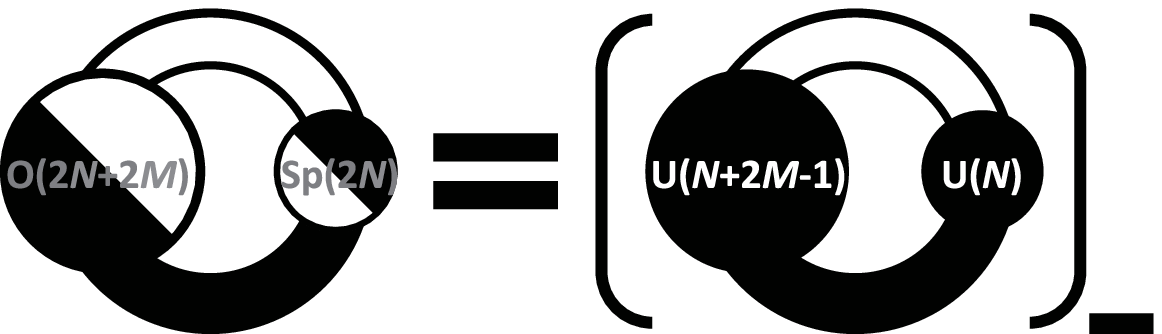}
\end{abstract}
\end{titlepage}

\tableofcontents

\renewcommand{\thefootnote}{\arabic{footnote}}
\setcounter{footnote}{0}

\section{Introduction and summary}\label{intro}
The ABJM theory \cite{ABJM} provides a rigorous formulation to study the M2-brane dynamics.
At the same time, it also provides a very profound mathematical structure.
In studying the partition function or one-point functions of the BPS Wilson loop, many interesting properties were found.
Most of these properties are shared with the superconformal Chern-Simons theories with a large number of supersymmetries ${\cal N}=6$ or ${\cal N}=5$ \cite{HLLLP2,ABJ}.

One interesting discovery is the hidden supergroup structure \cite{DT,MPtop}.
After using the localization techniques \cite{KWY}, the infinite-dimensional path integral in defining the partition function or the one-point functions of the theories is reduced to a finite-dimensional multiple integration.
These matrix models take the form of the Gaussian matrix model with two simultaneous deformations: the supergroup deformation and the trigonometric (or hyperbolic) deformation.
For the ABJM theory we replace the U$(N)$ gauge group of the Gaussian matrix model by the supergroup U$(N|N)$ and at the same time we change the rational functions by hyperbolic functions.
In terms of the hidden supergroups, other ${\cal N}=6$ or ${\cal N}=5$ superconformal theories are associated with U$(N_1|N_2)$ or OSp$(N_1|2N_2)$.

Another discovery is the Fermi gas formalism, which was first proposed in \cite{MP} for the original U$(N|N)$ ABJM theory.
It was found that the partition function of the U$(N|N+M)$ theory can also be rewritten into that of a Fermi gas system \cite{AHS,Ho1}\footnote{For an alternative formalism keeping the expression of the density matrix ${\widehat \rho}_{\text{U}(N|N)}$ in the U$(N|N+M)$ generalization, see \cite{MM}.}
\begin{align}
\biggl|\frac{Z_{\text{U}(N|N+M)}}{Z_{\text{U}(0|M)}}\biggr|
=\frac{1}{N!}\sum_{\sigma\in S_N}(-1)^\sigma
\int\frac{d^Nq}{(4\pi k)^N}
\prod_{i=1}^{N}\langle q_{\sigma(i)}|
\widehat\rho_{\text{U}(N|N+M)}|q_{i}\rangle,
\label{FG}
\end{align}
where the density matrix $\widehat\rho_{\text{U}(N|N+M)}$ relates a state $|q_{i}\rangle$ to its permutation $\langle q_{\sigma(i)}|$ which is accompanied by a sign factor $(-1)^\sigma$.
Using the position operator ${\widehat q}$ corresponding to the state $|q_i\rangle$ and the dual momentum operator ${\widehat p}$ obeying the canonical commutation relation $[{\widehat q},{\widehat p}]=i\hbar$ with $\hbar=2\pi k$, the density matrix is explicitly given by
\begin{align}
\widehat\rho_{\text{U}(N|N+M)}=\sqrt{V_M(\widehat q)}
\frac{1}{2\cosh\frac{\widehat p}{2}}
\sqrt{V_M(\widehat q)},
\label{denmatVM}
\end{align}
with \cite{Ho1,HoOk}
\begin{align}
V_M(q)=\frac{1}{e^{\frac{q}{2}}+(-1)^Me^{-\frac{q}{2}}}
\prod_{m=-\frac{M-1}{2}}^{\frac{M-1}{2}}\tanh\frac{q+2\pi im}{2k}.
\label{VM}
\end{align}
The Fermi gas formalism is not only beautiful but also practical.
In fact we can follow the systematic WKB (small $k$) analysis \cite{MP} to obtain the large $N$ expansion of the partition function.
Finally, the full large $N$ expansion was obtained \cite{HMMO,MM,HoOk}, based on the analysis in the Fermi gas formalism (small $k$ expansion \cite{MP,CM} and exact values for finite $k$ \cite{HMO1,PY,HMO2,HMO3}) together with the results from the 't Hooft expansion \cite{MPtop,DMP1,DMP2,FHM,KEK}.
See \cite{PTEP} for a review.

After establishing the result for the unitary supergroup, it is interesting to ask what happens if we replace the unitary supergroup by an orthosymplectic supergroup, whose physical interpretation is the introduction of an orientifold plane in the type IIB setup.
It was pointed out in \cite{MePu}\footnote{See \cite{Ok1} for a recent application.} that generally in studying the theories with orthosymplectic groups the projected density matrices introduced in \cite{HMO1}
\begin{align}
[\widehat\rho_{\text{U}(N|N+M)}]_\pm=\sqrt{V_M(\widehat q)}
\frac{\widehat\Pi_\pm}{2\cosh\frac{\widehat p}{2}}
\sqrt{V_M(\widehat q)},\quad
\widehat\Pi_\pm=\frac{1\pm \widehat R}{2},
\label{denmatVMproj}
\end{align}
play crucial roles, where $\widehat R$ is a reflection operator, ${\widehat R}|q\rangle =|-q\rangle$.
Owing to the large number of the supersymmetries, we hope that the non-perturbative effect of the orientifold plane can be clearly identified by studying the theories with orthosymplectic supergroups.
With this expectation, recently in \cite{MS1} the full large $N$ expansion of the OSp$(2N|2N)$ theory was studied with the even projected density matrix and a relation to the original ABJM U$(N|N)$ theory was found by doubling the orthosymplectic quiver in the sense of \cite{HoMo}.
Along these directions, interestingly it was found \cite{Ho2} that the density matrix for the OSp$(2N+1|2N)$ theory is identical to that for the ABJM U$(N|N)$ theory with the odd projection
\begin{align}
\widehat\rho_{\text{OSp}(2N+1|2N)}
=\bigl[\widehat\rho_{\text{U}(N|N)}\bigr]_-,
\end{align}
which allows us to study the OSp$(2N+1|2N)$ theory directly from the ABJM theory.
Subsequently this claim was generalized to the case of non-equal ranks \cite{MS2}\footnote{
The partition functions for the U$(N_1|N_2)$ and U$(N_2|N_1)$ theories are complex conjugate to each other, while the density matrices are identical \eqref{FG}.
Though, as a convention, we typically consider ${\widehat \rho}_{\text{U}(N_1|N_2)}$ with $N_2\ge N_1$, the schematic pattern may be clearer if we align the ranks of the unitary supergroups with those of the orthosymplectic supergroups.
}
\begin{align}
\widehat\rho_{\text{OSp}(2N+2M+1|2N)}
=\widehat\rho_{\text{OSp}(2N+1|2N+2M)}
=\bigl[\widehat\rho_{\text{U}(N|N+2M)}\bigr]_-.
\label{MSresult}
\end{align}
Related to these results, it was also observed in \cite{Ok2} that the values of the partition function for the OSp$(2N|2N)$ theory \cite{MS1} match with those for the U$(N|N+1)$ theory with the even projection for various integral values of $k$.
From this we naturally expect the relation
\begin{align}
\widehat\rho_{\text{OSp}(2N|2N)}
=\bigl[\widehat\rho_{\text{U}(N|N+1)}\bigr]_+,
\label{2N|2N}
\end{align}
though the proof has not been known.

\begin{figure}
\centering\includegraphics[scale=0.68]{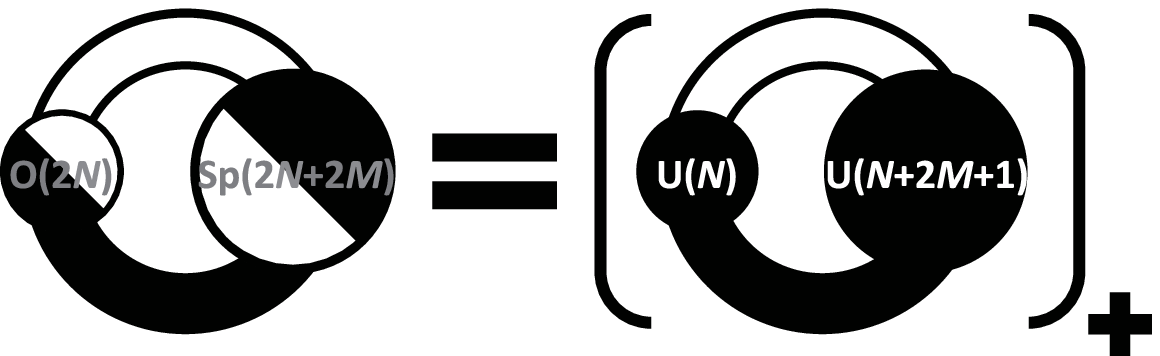}\qquad\includegraphics[scale=0.68]{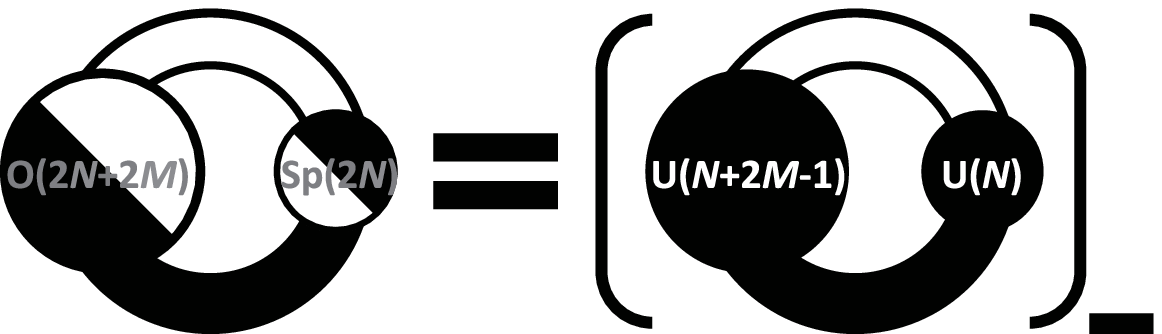}
\caption{Schematic relations between the density matrix for the theories with the orthosymplectic supergroups and those with the unitary supergroups.}
\label{osppic}
\end{figure}
In the first part of this paper, we shall fill this gap.
Namely, we first prove \eqref{2N|2N} by generalizing it to 
\begin{align}
\widehat\rho_{\text{OSp}(2N|2N+2M)}
=\bigl[\widehat\rho_{\text{U}(N|N+2M+1)}\bigr]_+,
\label{even}
\end{align}
with a non-negative integer $M\ge 0$.
It is natural to ask what happens to the case when the rank of the orthogonal bosonic subgroup is greater than that of the symplectic subgroup.
We answer this question by proving $(M\ge 1)$
\begin{align}
\widehat\rho_{\text{OSp}(2N+2M|2N)}
=\bigl[\widehat\rho_{\text{U}(N|N+2M-1)}\bigr]_-.
\label{odd}
\end{align}
The result is schematically summarized in figure \ref{osppic}.
Combined with the results \eqref{MSresult} from \cite{MS2}, we find an interesting pattern depicted in table \ref{relation}.
All of these relations completely reduce the study of the theories with orthosymplectic supergroups into that for the unitary supergroups with chiral projections.
\begin{table}[!ht]
\begin{align*}
&&
\widehat\rho_{\text{OSp}(2N+1|2N)}&=[\widehat\rho_{\text{U}(N|N)}]_-\\
\widehat\rho_{\text{OSp}(2N|2N)}&=[\widehat\rho_{\text{U}(N|N+1)}]_+&
\widehat\rho_{\text{OSp}(2N+2|2N)}&=[\widehat\rho_{\text{U}(N+1|N)}]_-\\
\widehat\rho_{\text{OSp}(2N+1|2N+2)}&=[\widehat\rho_{\text{U}(N|N+2)}]_-&
\widehat\rho_{\text{OSp}(2N+3|2N)}&=[\widehat\rho_{\text{U}(N+2|N)}]_-\\
\widehat\rho_{\text{OSp}(2N|2N+2)}&=[\widehat\rho_{\text{U}(N|N+3)}]_+&
\widehat\rho_{\text{OSp}(2N+4|2N)}&=[\widehat\rho_{\text{U}(N+3|N)}]_-\\
&\vdots&&\vdots\\
\widehat\rho_{\text{OSp}(2N|2N+k-2)}&
=[\widehat\rho_{\text{U}(N|N+k-1)}]_+&
\widehat\rho_{\text{OSp}(2N+k|2N)}&
=[\widehat\rho_{\text{U}(N+k-1|N)}]_-\\
\widehat\rho_{\text{OSp}(2N+1|2N+k)}&
=[\widehat\rho_{\text{U}(N|N+k)}]_-&
\widehat\rho_{\text{OSp}(2N+k+1|2N)}&
=[\widehat\rho_{\text{U}(N+k|N)}]_-
\end{align*}
\caption{Relations of the density matrices between the theory with orthosymplectic supergroups and that with unitary supergroups.}
\label{relation}
\end{table}

The key observation in our proof is to utilize the extra hyperbolic sine function $(2\sinh\frac{\nu}{k})^2$ appearing in the measure of the bosonic symplectic subgroup.
Although this factor was cumbersome in the Fermi gas formalism proposed in \cite{MS1}, owing to this factor we can naturally introduce a hyperbolic tangent function in the matrix elements, which is Fourier-dual to a hyperbolic cosecant function \cite{ADF,MN4}, reproducing the same factor in \eqref{VM} for odd $M$.

In \cite{MS2} only the Fermi gas system for the density matrices with the odd projection is assigned a physical meaning from the theories with orthosymplectic supergroups.
Here we have seen that both the odd and even chiral projections are physically relevant.

Note that the relations \eqref{even} and \eqref{odd} are consistent with the duality.
The duality for the theories with orthosymplectic supergroups \cite{ABJ}\footnote{Our convention of the level $k$ for the orthosymplectic theories is different from that in \cite{ABJ}: $k^\text{ABJ}=k^\text{here}/2$.}
\begin{align}
\text{OSp}(2N|2N+2M)&\Leftrightarrow\text{OSp}(2N|2N+2(k/2-M-1)),
\nonumber\\
\text{OSp}(2N+2M|2N)&\Leftrightarrow\text{OSp}(2N+2(k/2-M+1)|2N),
\end{align}
is translated to
\begin{align}
\text{U}(N|N+2M+1)&\Leftrightarrow\text{U}(N|N+2(k/2-M-1)+1),
\nonumber\\
\text{U}(N|N+2M-1)&\Leftrightarrow\text{U}(N|N+2(k/2-M+1)-1),
\end{align}
which is consistent with the duality for the theories with unitary supergroups.
In \cite{ABJ} the duality leads respectively to the constraint $0\le M\le k/2-1$ for the OSp$(2N|2N+2M)$ theories and $1\le M\le k/2$ for the OSp$(2N+2M|2N)$ theories.\footnote{Strictly speaking, although the constraint $0\le M\le k/2+1$ for the OSp$(2N+2M|2N)$ theories was proposed in \cite{ABJ}, in our analysis we find it more natural to exclude the $M=0$ case and the dual $M=k/2-1$ case in the OSp$(2N+2M|2N)$ theories.}

After establishing the relation between the theories with orthosymplectic supergroups and those with unitary supergroups with the chiral projections, in the second part we proceed to study the instanton effects \cite{BBS,DMP2} in the chirally projected $U(N|N+M)$ theories.
First of all we introduce the chemical potential $\mu$ dual to the particle number $N$ and switch from the Fermi gas partition function \eqref{FG} to the grand potential ${\widetilde J}(\mu)$ defined by \cite{MP,MM}
\begin{align}
e^{{\widetilde J}(\mu)}
=\sum_{N=0}^\infty e^{\mu N}\biggl|\frac{Z(N)}{Z(N=0)}\biggr|
=\det(1+e^\mu{\widehat\rho}),
\end{align}
where on the right-hand side we have combined the permutations among the $N$ integration variables in \eqref{FG} into the Fredholm determinant.
As its periodicity in $\mu\rightarrow\mu+2\pi i$ causes an oscillating behavior, it is reasonable to decompose the grand potential into the non-oscillating part $J(\mu)$ and the oscillations as
\begin{align}
e^{{\widetilde J}(\mu)}
=e^{J(\mu)}\biggl[1+\sum_{n\neq 0}e^{J(\mu+2\pi in)-J(\mu)}\biggr].
\end{align}
Hereafter the original grand potential ${\widetilde J}(\mu)$ is referred to as the {\it full} grand potential while the non-oscillating part $J(\mu)$ as the {\it modified} grand potential.
The instanton effects appear as the non-perturbative effects in the chemical potential ${\cal O}(e^{-\mu})$ in the large $\mu$ expansion of the modified grand potential $J(\mu)$.

To investigate the instanton effects in the chirally projected theories, we shall consider the two combinations $\Sigma(\mu)=J_+(\mu)+J_-(\mu)$ and $\Delta(\mu)=J_+(\mu)-J_-(\mu)$, instead of $J_\pm(\mu)$, the original modified grand potentials for the even and odd projected systems.
In the previous works \cite{MS1,MS2,Ok2} the modified grand potentials are studied for $k=1,2,3,4,5,6,8,12$ and $M=0,1,2,3,5$, and it was observed that there are two kinds of the instanton effects, the worldsheet instantons $(e^{-\frac{4m}{k}\mu})$ and the membrane instantons $(e^{-\ell\mu})$ $(m,\ell\in\mathbb{N})$.
Especially it was observed that the anti-symmetric combination $\Delta(\mu)$ contains only the membrane instantons $(e^{-\mu})$, which was also studied in full detail by the WKB expansion in \cite{Ok2}.

Our main interest in this paper is the instantons in the symmetric combination $\Sigma(\mu)$.
This part contains not only the membrane instantons but also the worldsheet instantons $(e^{-\frac{4}{k}\mu})$, which are non-perturbative in $k$ and thus cannot be analyzed in the WKB expansion.\footnote{
There also exist the bound states of these instantons which have the mixed exponents $(e^{-(\frac{4m}{k}+\ell)\mu})$.
It was observed in \cite{MS1,MS2}, however, that these effects are completely absorbed by the shift of the chemical potential $\mu$ into the effective chemical potential $\mu_\text{eff}$ \eqref{mueff} which is defined in the same way as in the ABJM theory without chiral projections \cite{HMO3}.
}
Instead, for this part we can utilize a trivial relation among the full grand potential for ${\widehat \rho}_{\text{U}(N|N+M)}$ and those for $[{\widehat \rho}_{\text{U}(N|N+M)}]_\pm$
\begin{align}
e^{{\widetilde J}(\mu)}
=e^{{\widetilde J}_+(\mu)}e^{{\widetilde J}_-(\mu)},
\label{trivial_infull}
\end{align}
which is, in turn, translated to a relation among $\Sigma(\mu)$, $\Delta(\mu)$ and the modified grand potential of the unprojected system $J(\mu)$.
Note that this relation also implies that the perturbative part and the membrane instantons in $\Sigma(\mu)$ completely coincide with those in the system without chiral projections.

In this paper we study this relation carefully.
By taking the explicit results for $\Delta(\mu)$ into account, we finally identify a simple relation without the oscillatory behavior,
\begin{align}
J(\mu)
&=\Sigma(\mu)+\log\biggl(1+2\sum_{n=1}^\infty(-1)^n
e^{\frac{1}{2}\Sigma(\mu+2\pi in)+\frac{1}{2}\Sigma(\mu-2\pi in)
-\Sigma(\mu)}
\biggr).
\label{JSigma}
\end{align}
The result is derived in a similar way as for the duplicate quivers \cite{HoMo}.
With this relation the worldsheet instanton in $\Sigma(\mu)$ is completely solved in terms of the result of the ABJM theory.

Surprisingly, we find that the worldsheet instanton in the chirally projected theories fits well with the Gopakumar-Vafa formula for the topological string partition function \cite{GV2}.
For the worldsheet instanton in the unprojected $U(N|N+M)$ theories, it was already known that the same formula with the topological invariants on local $\mathbb{P}^1\times\mathbb{P}^1$ works \cite{MPtop}.
It is still non-trivial that the formula applies for the chirally projected theories since the relation \eqref{JSigma} induces new instanton effects non-linearly.
Nevertheless, with \eqref{JSigma} we can check it and identify the Gopakumar-Vafa invariants $n^{\bm d}_g$ for the chirally projected theories up to the seventh instanton as in tables \ref{GV12345} and \ref{GV67}.
At present the interpretation of the invariants $n^{\bm d}_g$ is unclear.
We shall briefly argue this point in section \ref{discussion}.

The organization of this paper is as follows.
In section \ref{chiral}, we prove the relations \eqref{even} and \eqref{odd}.
After reducing the question for the theories with orthosymplectic supergroups into that for the theories with unitary supergroups and the chiral projections, in section \ref{ws} we study the grand potential for the latter theories.
Finally we conclude with discussions in section \ref{discussion}.

\section{Chiral projections}\label{chiral}
In this section, we prove that the Fermi gas system for the OSp$(2N|2N+2M)$ theory with $0\le M\le k/2-1$ matches with that for the U$(N|N+2M+1)$ theory with the even projection, while the Fermi gas system for the OSp$(2N+2M|2N)$ theory with $1\le M\le k/2$ is that for the U$(N|N+2M-1)$ theory with the odd projection.
Namely, we study the partition function of the theory with the orthosymplectic supergroup carefully and finally find that the partition function is rewritten into
\begin{align}
Z_\pm(N)
=\frac{1}{N!}\int\frac{d^Nq}{(4\pi k)^N}
\prod_{i=1}^NV(q_i)\pi_\pm(q_i)
\prod_{i<j}^N
\biggl(\tanh\frac{q_i-q_j}{2k}\tanh\frac{q_i+q_j}{2k}\biggr)^2,
\label{pumi}
\end{align}
with some function $V(q)$ and $\pi_\pm(q)$ given by
\begin{align}
\pi_+(q)=\frac{\cosh^2\frac{q}{2k}}{\cosh\frac{q}{k}},\quad
\pi_-(q)=\frac{\sinh^2\frac{q}{2k}}{\cosh\frac{q}{k}},\quad
\pi_+(q)+\pi_-(q)=1.
\end{align}
The right-hand side of \eqref{pumi} is known to be written as the partition function of the $N$ particle ideal Fermi gas system \eqref{FG} with the density matrix
\begin{align}
\widehat\rho=\sqrt{V(\widehat q)}
\frac{\widehat\Pi_\pm}{2\cosh\frac{\widehat p}{2}}
\sqrt{V(\widehat q)},
\end{align}
which coincide with the chirally projected density matrix for the $\text{U}(N|N+M)$ theory \eqref{denmatVMproj} if $V(q)=V_M(q)$.
The proof goes mostly in parallel with \cite{MS2} except only a few important novelties.

\subsection{Even projection}\label{seceven}
In this subsection we provide the proof of \eqref{even}.
After the localization technique, the partition function is \cite{KWY,GHN}
\begin{align}
Z_{\text{OSp}(2N|2N+2M)}=\frac{1}{N!(N+M)!}\int\frac{d^N\mu}{(4\pi k)^N}
\frac{d^{N+M}\nu}{(4\pi k)^{N+M}}
e^{\frac{i}{4\pi k}(\sum_{i=1}^N\mu_i^2-\sum_{k=1}^{N+M}\nu_k^2)}
\times Z_\text{1-loop},
\end{align}
with the one-loop determinant factor
\begin{align}
&Z_\text{1-loop}\nonumber \\
&=\frac{\prod_{i<j}^N(2\sinh\frac{\mu_i-\mu_j}{2k})^2
(2\sinh\frac{\mu_i+\mu_j}{2k})^2
\prod_{k<l}^{N+M}
(2\sinh\frac{\nu_k-\nu_l}{2k})^2
(2\sinh\frac{\nu_k+\nu_l}{2k})^2
\prod_{k=1}^{N+M}(2\sinh\frac{\nu_k}{k})^2}
{\prod_{i=1}^N\prod_{k=1}^{N+M}(2\cosh\frac{\mu_i-\nu_k}{2k})^2
(2\cosh\frac{\mu_i+\nu_k}{2k})^2}.
\end{align}
As in \cite{MM,MS2} our starting point is the Cauchy-Vandermonde determinant
\begin{align}
&\det\begin{pmatrix}
\Bigl[\frac{1}{(2\cosh\frac{\mu_i-\nu_k}{2k})
(2\cosh\frac{\mu_i+\nu_k}{2k})}\Bigr]_{(i,k)\in Z_N\times Z_{N+M}}\\
\Bigl[\frac{\sinh\frac{m\nu_k}{k}}{\sinh\frac{\nu_k}{k}}\Bigr]
_{(m,k)\in Z_M\times Z_{N+M}}
\end{pmatrix}=(-1)^{MN+\frac{1}{2}M(M-1)}\nonumber\\
&\quad\times\frac{\prod_{i<j}^N(2\sinh\frac{\mu_i-\mu_j}{2k})
(2\sinh\frac{\mu_i+\mu_j}{2k})
\prod_{k<l}^{N+M}(2\sinh\frac{\nu_k-\nu_l}{2k})
(2\sinh\frac{\nu_k+\nu_l}{2k})}
{\prod_{i=1}^M\prod_{k=1}^{N+M}(2\cosh\frac{\mu_i-\nu_k}{2k})
(2\cosh\frac{\mu_i+\nu_k}{2k})},
\label{CauchyVdM}
\end{align}
where we have denoted the integer set spanned by each index with the notation $Z_L=\{1,2,\cdots,L\}$.
After multiplying $2\sinh\frac{\nu_k}{k}$ to each column, we obtain the following determinant expression for $Z_\text{1-loop}$
\begin{align}
Z_\text{1-loop}=
\det\begin{pmatrix}
\Bigl[\frac{2\sinh\frac{\nu_k}{k}}
{(2\cosh\frac{\mu_i-\nu_k}{2k})(2\cosh\frac{\mu_i+\nu_k}{2k})}\Bigr]
_{(i,k)\in Z_N\times Z_{N+M}}\\
\Bigl[2\sinh\frac{m\nu_k}{k}\Bigr]_{(m,k)\in Z_M\times Z_{N+M}}
\end{pmatrix}^2.
\label{1-loop_in_det}
\end{align}

To proceed, it is helpful to introduce the canonical position and momentum operators ${\widehat q}$ and ${\widehat p}$ obeying the commutation relation $[{\widehat q},{\widehat p}]=i\hbar$ with $\hbar=2\pi k$. 
We shall also introduce the states $|m\rrb$ to abbreviate the matrix components in the lower block of \eqref{1-loop_in_det}
\begin{align}
2\sinh\frac{m\nu_k}{k}=\llb m|\nu_k\rangle=\langle\nu_k|m\rrb,
\label{[]}
\end{align}
where $|\nu\rangle$ is the coordinate eigenstate normalized as $\langle\mu|\nu\rangle=2\pi\delta(\mu-\nu)$.
For the upper block, it is crucial to express the matrix component as\footnote{We assume that the singularity at $\widehat p=0$ is cancelled by the projections and hence harmless.}
\begin{align}
&\frac{2\sinh\frac{\nu_k}{k}}{(2\cosh\frac{\mu_i-\nu_k}{2k})
(2\cosh\frac{\mu_i+\nu_k}{2k})}
=-\frac{1}{2}\Bigl(\tanh\frac{\mu_i-\nu_k}{2k}
-\tanh\frac{\mu_i+\nu_k}{2k}\Bigr)\nonumber\\
&\quad=2ik\langle\mu_i|\frac{1}{2\sinh\frac{\widehat p}{2}}
\widehat\Pi_-|\nu_k\rangle
=-2ik\langle\nu_k|\frac{1}{2\sinh\frac{\widehat p}{2}} \widehat\Pi_+|\mu_i\rangle,
\label{tanh-tanh}
\end{align}
using the Fourier duality between hyperbolic tangent functions and hyperbolic cosecant functions \cite{ADF,MN4}.
This is the most important point in our proof.

After noticing this structure the rest of the computation is quite parallel to \cite{MS2}: substitute $Z_\text{1-loop}$ \eqref{1-loop_in_det} with \eqref{[]} and \eqref{tanh-tanh}; include the Fresnel factors $e^{\frac{i}{4\pi k}\mu_i^2}$ and $e^{-\frac{i}{4\pi k}\nu_k^2}$ into the brackets; trivialize the first determinant by renaming the indices of $\nu_k$; perform the similarity transformation
\begin{align}
1=\int\frac{dq}{2\pi}
e^{-\frac{i}{2\hbar}{\widehat p}^2}|q\rangle
\langle q|e^{\frac{i}{2\hbar}{\widehat p}^2},
\end{align}
to the integration variables.
After these steps we obtain the expression
\begin{align}
&Z_{\text{OSp}(2N|2N+2M)}=\frac{1}{N!}\int\frac{d^N\mu}{(4\pi k)^N}
\frac{d^{N+M}\nu}{(4\pi k)^{N+M}}\nonumber\\
&\quad\times\prod_{i=1}^N2ik\langle\mu_i|e^{\frac{i}{2\hbar}\widehat p^2}
e^{\frac{i}{2\hbar}\widehat q^2}
\frac{1}{2\sinh\frac{\widehat p}{2}}\widehat\Pi_-
e^{-\frac{i}{2\hbar}\widehat q^2}
e^{-\frac{i}{2\hbar}\widehat p^2}|\nu_i\rangle
\prod_{m=1}^M\llb m|e^{-\frac{i}{2\hbar}\widehat q^2}
e^{-\frac{i}{2\hbar}\widehat p^2}|\nu_{N+m}\rangle\nonumber\\
&\quad\times\det\begin{pmatrix}
\Bigl[-2ik\langle\nu_k|\frac{1}{2\sinh\frac{\widehat p}{2}}
\widehat\Pi_+|\mu_j\rangle\Bigr]
_{(k,j)\in Z_{N+M}\times Z_N}&
\Bigl[\langle\nu_k|e^{\frac{i}{2\hbar}\widehat p^2}|n\rrb\Bigr]
_{(k,n)\in Z_{N+M}\times Z_M}\end{pmatrix}.
\label{Zeven2}
\end{align}
As in \cite{MS2} the elements in the two products in front of the determinant become the delta functions and we can perform the integrations.
We shall see this explicitly in the following.

Let us first simplify the determinant.
Using
\begin{align}
\langle\nu_k|e^{\frac{i}{2\hbar}\widehat p^2}|n\rrb
=e^{-\frac{i}{2\hbar}(2\pi n)^2}\langle\nu_k|n\rrb,
\end{align}
the determinant reduces to
\begin{align}
&\det\begin{pmatrix}
\Bigl[
-2ik\langle\nu_k|\frac{1}{2\sinh\frac{\widehat p}{2}}
\widehat\Pi_+|\mu_j\rangle
\Bigr]_{Z_{N+M}\times Z_N}
&
\Bigl[
\langle\nu_k|e^{\frac{i}{2\hbar}\widehat p^2}|n\rrb
\Bigr]_{Z_{N+M}\times Z_M}
\end{pmatrix}\nonumber\\
&=e^{-\frac{i}{12\hbar}(2\pi)^2M(2M+1)(M+1)}
\det\begin{pmatrix}
\Bigl[
-2ik\langle\nu_k|\frac{1}{2\sinh\frac{\widehat p}{2}}
\widehat\Pi_+|\mu_j\rangle
\Bigr]_{Z_{N+M}\times Z_N}
&
\Bigl[
\langle\nu_k|n\rrb
\Bigr]_{Z_{N+M}\times Z_M}
\end{pmatrix},
\label{evendet}
\end{align}
which is an odd function of $\nu_k$.

The two products in the second line of \eqref{Zeven2} can be formally computed as\footnote{
In the second line in \eqref{deltaIm} we assume the following deformation of the integration contour for $\nu_{N+m}$
\begin{align}
(-\infty,\infty)\rightarrow
(-\infty-2\pi im,-2\pi im)
\sqcup [-2\pi im,2\pi im]
\sqcup (2\pi im,\infty+2\pi im)
\end{align}
so that the new contour contains the supports of the delta functions.
We can show that such a deformation is allowed if and only if $M<k/2$, following the argument in \cite{MS2}.
}
\begin{align}
&2ik\langle\mu_i|e^{\frac{i}{2\hbar}\widehat p^2}
e^{\frac{i}{2\hbar}\widehat q^2}
\frac{1}{2\sinh\frac{\widehat p}{2}}\widehat\Pi_-
e^{-\frac{i}{2\hbar}\widehat q^2}
e^{-\frac{i}{2\hbar}\widehat p^2}|\nu_i\rangle
=\frac{2\pi k}{i}\frac{1}{2\sinh\frac{\mu_i}{2}}
(\delta(\mu_i-\nu_i)-\delta(\mu_i+\nu_i)),
\nonumber\\
&\llb m|e^{-\frac{i}{2\hbar}\widehat q^2}
e^{-\frac{i}{2\hbar}\widehat p^2}|\nu_{N+m}\rangle
=\frac{2\pi k}{\sqrt{ik}}e^{-\frac{i}{2\hbar}(2\pi m)^2}
(\delta(\nu_{N+m}+2\pi im)-\delta(\nu_{N+m}-2\pi im)).
\label{deltaIm}
\end{align}
Since the rest of the integrand \eqref{evendet} is an odd function of $\nu_k$, we can drop one of the two delta functions as in \cite{MS2},
\begin{align}
&2ik\langle\mu_i|e^{\frac{i}{2\hbar}\widehat p^2}
e^{\frac{i}{2\hbar}\widehat q^2}
\frac{1}{2\sinh\frac{\widehat p}{2}}\widehat\Pi_-
e^{-\frac{i}{2\hbar}\widehat q^2}
e^{-\frac{i}{2\hbar}\widehat p^2}|\nu_i\rangle
\to\frac{4\pi k}{i}\frac{1}{2\sinh\frac{\mu_i}{2}}
\delta(\nu_i-\mu_i),
\nonumber\\
&\llb m|e^{-\frac{i}{2\hbar}\widehat q^2}
e^{-\frac{i}{2\hbar}\widehat p^2}|\nu_{N+m}\rangle
\to-\frac{4\pi k}{\sqrt{ik}}e^{-\frac{i}{2\hbar}(2\pi m)^2}
\delta(\nu_{N+m}-2\pi im).
\label{evendelta}
\end{align}

After reducing the two matrix elements into the delta functions we can perform the $\nu_k$ integrations by simple substitutions, leaving only a single determinant in \eqref{Zeven2}.
Reverting it into the products \eqref{CauchyVdM} and separating the $N$-independent factors, we finally find that the partition function \eqref{Zeven2} is given by $Z_+(N)$ \eqref{pumi} as
\begin{align}
\frac{(-1)^{MN}i^NZ_{\text{OSp}(2N|2N+2M)}}{Z_{\text{OSp}(0|2M)}}
=Z_+(N),
\label{resulteven}
\end{align}
with $Z_{\text{OSp}(0|2M)}$ being an $N$-independent factor
\begin{align}
Z_{\text{OSp}(0|2M)}
&=(-1)^{\frac{1}{2}M(M+1)}e^{-\frac{\pi i}{3k}M(2M+1)(M+1)}
(ik)^{-\frac{M}{2}}\nonumber\\
&\quad\times
\prod_{m<n}^M4\sinh\frac{\rho_m-\rho_n}{2k}\sinh\frac{\rho_m+\rho_n}{2k}
\prod_{m=1}^M2\sinh\frac{\rho_m}{k},
\end{align}
(which is non-vanishing for $0\le M\le k/2-1$) and $V(\mu)$ given by
\begin{align}
V(\mu)=\frac{1}{2\sinh\frac{\mu}{2}}\tanh\frac{\mu}{2k}
\prod_{m=1}^M\tanh\frac{\mu-\rho_m}{2k}\tanh\frac{\mu+\rho_m}{2k},
\end{align}
with $\rho_m=2\pi im$.
Comparing with the expression \eqref{VM} we find that $V(\mu)=V_{2M+1}(\mu)$.
In this way we have proved that the Fermi gas system for the OSp$(2N|2N+2M)$ theory is identical to that for the U$(N|N+2M+1)$ theory with the even chiral projection.

\subsection{Odd projection}
In the previous subsection, we have seen that the density matrix for the theory with orthosymplectic supergroups where the rank of the bosonic symplectic subgroup is greater than or equal to that of the orthogonal subgroup is related to that for the unitary supergroups.
Here we shall turn to the opposite case: OSp$(2N+2M|2N)$ with $1\le M\le k/2$.

The partition function is \cite{GHN,KWY}
\begin{align}
&Z_{\text{OSp}(2N+2M|2N)}
=\frac{1}{N!(N+M)!}\int\frac{d^{N+M}\mu}{(4\pi k)^{N+M}}
\frac{d^{N}\nu}{(4\pi k)^{N}}
e^{\frac{i}{4\pi k}(\sum_{i=1}^{N+M}\mu_i^2-\sum_{k=1}^{N}\nu_k^2)}
\nonumber\\
&\times\frac{\prod_{i<j}^{N+M}(2\sinh\frac{\mu_i-\mu_j}{2k})^2
(2\sinh\frac{\mu_i+\mu_j}{2k})^2
\prod_{k<l}^{N}
(2\sinh\frac{\nu_k-\nu_l}{2k})^2
(2\sinh\frac{\nu_k+\nu_l}{2k})^2
\prod_{k=1}^{N}(2\sinh\frac{\nu_k}{k})^2}
{\prod_{i=1}^{N+M}\prod_{k=1}^{N}(2\cosh\frac{\mu_i-\nu_k}{2k})^2
(2\cosh\frac{\mu_i+\nu_k}{2k})^2}.
\end{align}
Since we can add one row by a multiple of another row in the determinant without changing its value, we can express the Cauchy-Vandermonde determinant \eqref{CauchyVdM} by
\begin{align}
&\det\begin{pmatrix}
\Bigl[\frac{1}{(2\cosh\frac{\mu_i-\nu_k}{2k})
(2\cosh\frac{\mu_i+\nu_k}{2k})}\Bigr]_{(i,k)\in Z_{N+M}\times Z_{N}}&
\Bigl[2\cosh\frac{(m-1)\mu_i}{k}\Bigr]
_{(m,k)\in Z_{N+M}\times Z_{M}}
\end{pmatrix}\nonumber\\
&=2(-1)^{MN+\frac{1}{2}M(M-1)}
\frac{\prod_{i<j}^{N+M}(2\sinh\frac{\mu_i-\mu_j}{2k})
(2\sinh\frac{\mu_i+\mu_j}{2k})
\prod_{k<l}^{N}(2\sinh\frac{\nu_k-\nu_l}{2k})
(2\sinh\frac{\nu_k+\nu_l}{2k})}
{\prod_{i=1}^{N+M}\prod_{k=1}^{N}(2\cosh\frac{\mu_i-\nu_k}{2k})
(2\cosh\frac{\mu_i+\nu_k}{2k})}.
\end{align}
After multiplying $2\sinh\frac{\nu_k}{k}$ to each column, the left block in the determinant can be expressed as the previous case \eqref{tanh-tanh}.
For the right block we introduce the states $|m\rrp$ by
\begin{align}
2\cosh\frac{(m-1)\mu}{k}=\llp m|\mu\rangle=\langle\mu|m\rrp.
\label{()}
\end{align}
Then, through the same steps as in the last subsection, we can rewrite the partition function as
\begin{align}
&Z_{\text{OSp}(2N+2M|2N)}=\frac{1}{N!}
\int\frac{d^N\nu}{(4\pi k)^N}\frac{d^{N+M}\mu}{(4\pi k)^{N+M}}
\nonumber\\
&\quad\times\prod_{k=1}^N2ik\langle\mu_k|e^{\frac{i}{2\hbar}\widehat p^2}
e^{\frac{i}{2\hbar}\widehat q^2}
\frac{1}{2\sinh\frac{\widehat p}{2}}\widehat\Pi_-
e^{-\frac{i}{2\hbar}\widehat q^2}e^{-\frac{i}{2\hbar}\widehat p^2}
|\nu_k\rangle
\prod_{m=1}^M\langle\mu_{N+m}|
e^{\frac{i}{2\hbar}\widehat p^2}e^{\frac{i}{2\hbar}\widehat q^2}
|m\rrp
\nonumber\\&\quad\times
\det\begin{pmatrix}
\Bigl[-2ik\langle\nu_k|\frac{1}{2\sinh\frac{\widehat p}{2}}
\widehat\Pi_+|\mu_j\rangle\Bigr]_{(k,j)\in Z_{N}\times Z_{N+M}}\\
\Bigl[\llp n|e^{-\frac{i}{2\hbar}\widehat p^2}|\mu_j\rangle\Bigr]
_{(n,j)\in Z_{M}\times Z_{N+M}}
\end{pmatrix},
\end{align}
where the first two factors can be replaced with the delta functions
\begin{align}
&2ik\langle\mu_k|e^{\frac{i}{2\hbar}\widehat p^2}
e^{\frac{i}{2\hbar}\widehat q^2}
\frac{1}{2\sinh\frac{\widehat p}{2}}\widehat\Pi_-
e^{-\frac{i}{2\hbar}\widehat q^2}e^{-\frac{i}{2\hbar}\widehat p^2}
|\nu_k\rangle
\to\frac{4\pi k}{i}\frac{1}{2\sinh\frac{\nu_k}{2}}\delta(\mu_k-\nu_k),
\nonumber\\
&\langle\mu_{N+m}|
e^{\frac{i}{2\hbar}\widehat p^2}e^{\frac{i}{2\hbar}\widehat q^2}|m\rrp
\to\frac{4\pi k}{\sqrt{-ik}}e^{\frac{i}{2\hbar}(2\pi)^2(m-1)^2}
\delta(\mu_{N+m}-2\pi i(m-1)).
\end{align}

After performing the integration over $\mu_i$, this time we obtain
\begin{align}
\frac{i^N(-1)^{MN}Z_{\text{OSp}(2N+2M|2N)}}{Z_{\text{OSp}(2M|0)}}
=Z_-(N),
\label{resultodd}
\end{align}
with
\begin{align}
Z_{\text{OSp}(2M|0)}=2(-1)^{\frac{1}{2}M(M-1)}(-ik)^{-\frac{M}{2}}
e^{\frac{\pi i}{3k}M(2M-1)(M-1)}
\prod_{m<n}^M4\sinh\frac{\rho_{m-1}-\rho_{n-1}}{2k}
\sinh\frac{\rho_{m-1}+\rho_{n-1}}{2k},
\end{align}
and
\begin{align}
V(\nu)=\frac{1}{2\sinh\frac{\nu}{2}}\Bigl(\tanh\frac{\nu}{2k}\Bigr)^{-1}
\prod_{m=1}^M
\tanh\frac{\nu-\rho_{m-1}}{2k}\tanh\frac{\nu+\rho_{m-1}}{2k}.
\end{align}
Hence again \eqref{resultodd} is nothing but the Fermi gas formalism for the $U(N|N+2M-1)$ theory with the odd chiral projection.

\section{Worldsheet instantons}\label{ws}
In the previous section, combined with the result from \cite{MS1}, we have found that the matrix model associated with any orthosymplectic supergroup is directly related to that associated with the corresponding unitary supergroup and the chiral projection.
This means that we can study the OSp$(N_1|2N_2)$ theories with completely general $(N_1,N_2)$ from the $\text{U}(N|N+M)$ theories with the chiral projections.

Here we proceed to study the grand potentials for the latter theories.
First we define the following symmetric and anti-symmetric combination
\begin{align}
\Sigma(\mu)=J_+(\mu)+J_-(\mu),\quad
\Delta(\mu)=J_+(\mu)-J_-(\mu),\quad
\end{align}
where $J_\pm(\mu)$ are respectively the modified grand potential for the density matrix $[{\widehat \rho}_{\text{U}(N|N+M)}]_\pm$.
The large $\mu$ expansion of $\Sigma(\mu)$ and $\Delta(\mu)$ was studied previously in \cite{Ok2,MS2}, and was observed to have the following structures
\begin{align}
\Sigma(\mu)&=\frac{C}{3}\mu_\text{eff}^3+B\mu_\text{eff}+A
+\sum_{m=1}^\infty s_m e^{-\frac{4m}{k}\mu_\text{eff}}
+\sum_{\ell=1}^\infty
({\widetilde b}_\ell\mu_\text{eff}+{\widetilde c}_\ell)
e^{-2\ell\mu_\text{eff}},\nonumber\\
\Delta(\mu)&=\frac{\mu}{2}+A^\prime
+\sum_{\ell=1}^\infty r_\ell e^{-\ell\mu},
\label{schem}
\end{align}
with
\begin{align}
C=\frac{2}{\pi^2k},
\label{CABJM}
\end{align}
and $(B,A,s_m,{\widetilde b}_\ell,{\widetilde c}_\ell,A^\prime,r_\ell)$ being some constants depending on $k$ and $M$.
Here $\mu_\text{eff}$ is the effective chemical potential defined in the $\text{U}(N|N+M)$ theory without chiral projections \cite{HMO3,MM}
\begin{align}
\mu_\text{eff}=\mu+\frac{1}{C}\sum_{\ell=1}^\infty a_\ell e^{-2\ell\mu},
\label{mueff}
\end{align}
with some constants $a_\ell$.
The possible bound states of the worldsheet instantons and the membrane instantons $e^{-(\frac{4m}{k}+\ell)\mu}$ are observed to be absorbed into the worldsheet instantons $e^{-\frac{4m}{k}\mu_\text{eff}}$ through the non-perturbative deviations in $\mu_\text{eff}$ \eqref{mueff}.
Note that $\Sigma(\mu)$ and the modified grand potential without chiral projections $J(\mu)$ have the same schematic structure: they have completely the same perturbative part and the membrane instantons; they are different only in the coefficients of the worldsheet instantons.
Several coefficients of the half membrane instantons $r_\ell$ $(1\le \ell\le 7)$ in $\Delta(\mu)$ are determined as explicit functions of $(k,M)$ through the extrapolation of the WKB small $k$ expansion in \cite{Ok2}.

Below we shall study the worldsheet instantons in the symmetric combination $\Sigma(\mu)$.
With the help of the schematic expression \eqref{schem} of the anti-symmetric part $\Delta(\mu)$, we extract from \eqref{trivial_infull} a new relation \eqref{JSigma} between $\Sigma(\mu)$ and $J(\mu)$.
Using it we further compute the explicit expression of the first few instanton coefficients and find that they fit with the Gopakumar-Vafa formula of the topological string theory, with the Gopakumar-Vafa invariants identified as in tables \ref{diagGV}, \ref{GV12345} and \ref{GV67}.

\subsection{Worldsheet instantons for chiral projections}
From the fundamental properties of the projection operators ${\widehat\Pi}_++{\widehat\Pi}_-=1$ and ${\widehat \Pi}_+{\widehat \Pi}_-=0$,
it follows that
\begin{align}
\det(1+e^\mu{\widehat \rho}_{\text{U}(N|N+M)})
=\det(1+e^\mu[{\widehat \rho}_{\text{U}(N|N+M)}]_+)\det(1+e^\mu[{\widehat \rho}_{\text{U}(N|N+M)}]_-).
\end{align}
This is translated into the relations among the full grand potentials \eqref{trivial_infull} or
\begin{align}
\sum_{n\in\mathbb{Z}}e^{J(\mu+2\pi in)}=
\sum_{n_+\in\mathbb{Z}}e^{J_+(\mu+2\pi in_+)}
\sum_{n_-\in\mathbb{Z}}e^{J_-(\mu+2\pi in_-)}.
\label{totaltrivial}
\end{align}
At first sight it seems difficult to extract a relation between the modified grand potentials as the equation contains the oscillating terms both for $J(\mu)$ and $J_\pm(\mu)$.
As we see below, however, using the schematic expression of the anti-symmetric part $\Delta(\mu)$ \eqref{schem} we obtain a more refined relation \eqref{JSigma} where the oscillations for $J(\mu)$ are absent.

We rewrite the relation \eqref{totaltrivial} in terms of $\Sigma(\mu)$ and $\Delta(\mu)$
\begin{align}
\sum_{n\in\mathbb{Z}}e^{J(\mu+2\pi in)}
=\sum_{n_+\in\mathbb{Z}}e^{\frac{\Sigma+\Delta}{2}(\mu+2\pi in_+)}
\sum_{n_-\in\mathbb{Z}}e^{\frac{\Sigma-\Delta}{2}(\mu+2\pi in_-)}.
\end{align}
As $\Delta(\mu)$ in \eqref{schem} has the following simple quasi-periodicity
\begin{align}
\Delta(\mu+2\pi in)=\Delta(\mu)+\pi in,
\end{align}
the relation reduces to
\begin{align}
\sum_{n\in\mathbb{Z}}e^{J(\mu+2\pi in)}
=\sum_{n_+\in\mathbb{Z}}\sum_{n_-\in\mathbb{Z}}
i^{n_+-n_-}e^{(\Sigma(\mu+2\pi in_+)+\Sigma(\mu+2\pi in_-))/2}.
\end{align}
Here we notice that the terms with odd $n_+-n_-$ do not contribute, since each of those terms is always accompanied with the one with the opposite sign
\begin{align}
i^{n_+-n_-}
e^{\frac{1}{2}\Sigma(\mu+2\pi in_+)
+\frac{1}{2}\Sigma(\mu+2\pi in_-)}
+i^{n_--n_+}
e^{\frac{1}{2}\Sigma(\mu+2\pi in_-)
+\frac{1}{2}\Sigma(\mu+2\pi in_+)}
=0.
\end{align}
Then, \eqref{totaltrivial} can be rewritten as
\begin{align}
\sum_{n\in\mathbb{Z}}e^{J(\mu+2\pi in)}
=\sum_{n_+-n_-\in 2{\mathbb Z}}(-1)^{\frac{n_+-n_-}{2}}
e^{\frac{1}{2}\Sigma(\mu+2\pi in_+)
+\frac{1}{2}\Sigma(\mu+2\pi in_-)}.
\label{noodd}
\end{align}
This relation can be solved by
\begin{align}
e^{J(\mu)}=\sum_{n\in\mathbb{Z}}(-1)^n
e^{\frac{1}{2}\Sigma(\mu+2\pi in)
+\frac{1}{2}\Sigma(\mu-2\pi in)},
\label{solution}
\end{align}
which can be rewritten as \eqref{JSigma}.
As in \cite{HoMo} we can easily check that \eqref{solution} correctly reproduces \eqref{noodd} and contains no oscillations.

\subsection{Gopakumar-Vafa invariants}
To obtain the direct relation between the worldsheet instantons in $\Sigma(\mu)$ and those in $J(\mu)$, let us compute the exponent inside the logarithm in \eqref{JSigma} using the schematic expression of $\Sigma(\mu)$ \eqref{schem}.
Note that, since the effective chemical potential $\mu_\text{eff}$ is related to the original one $\mu$ by \eqref{mueff}, shifting $\mu$ by $\pm 2\pi in$ directly means shifting $\mu_\text{eff}$.
Hence, in the following we shall regard $\Sigma(\mu)$ in \eqref{schem} as a function of $\mu_\text{eff}$ and, with a slight abuse of notation, denote the same function as $\Sigma(\mu_\text{eff})$.
Then, most of the contributions in the exponent in \eqref{JSigma} including the non-perturbative membrane instantons cancel and we are left with the $C$ term and the worldsheet instantons as
\begin{align}
\frac{1}{2}\Sigma(\mu_\text{eff}+2\pi in)
+\frac{1}{2}\Sigma(\mu_\text{eff}-2\pi in)-\Sigma(\mu_\text{eff})
=-\frac{8n^2}{k}\mu_\text{eff}
-2\sum_{m=1}^\infty s_m\sin^2\frac{4\pi mn}{k}
e^{-\frac{4m\mu_\text{eff}}{k}},
\end{align}
where we have used the explicit value of $C$ \eqref{CABJM}.
Hence we obtain an equation containing only the worldsheet instantons
\begin{align}
J^\text{WS}(\mu_\text{eff})=\sum_{m=1}^\infty s_mz_\text{eff}^m
+\log\biggl(1+2\sum_{n=1}^\infty(-1)^nz_\text{eff}^{2n^2}
\prod_{m=1}^\infty
e^{-2s_m\sin^2\frac{4\pi mn}{k}z_\text{eff}^m}\biggr),
\label{WSeq}
\end{align}
where $z_\text{eff}=e^{-\frac{4}{k}\mu_\text{eff}}$ and $J^\text{WS}(\mu_\text{eff})$ is the worldsheet instantons in the $U(N|N+M)$ theory without chiral projections \cite{HMMO,HoOk}
\begin{align}
J^\text{WS}(\mu_\text{eff})=\sum_{m=1}^\infty d_mz_\text{eff}^m.
\end{align}
We can solve \eqref{WSeq} inversely order by order in $z_\text{eff}$ as
\begin{align}
&s_1=d_1,\quad
s_2=d_2+2,\quad
s_3=d_3-4d_1\sin^2\frac{4\pi}{k},\nonumber\\
&s_4=d_4-4d_2\sin^2\frac{8\pi}{k}
+4d_1^2\sin^4\frac{4\pi}{k}+2-8\sin^2\frac{8\pi}{k},\quad\cdots.
\end{align}
Then, we find that the functional form of $s_m$ reproduces all the numerical fitting in \cite{Ok2,MS2}.
More surprisingly, this functional form fit well with the Gopakumar-Vafa formula, which is not guaranteed from the beginning,
\begin{align}
s_m(k,M)=(-1)^m\sum_{nd=m}\frac{1}{n}
\sigma_{{d}}\Bigl(\frac{k}{n},M\Bigr),\quad
\sigma_d(k,M)=\sum_{d_1+d_2=d}e^{2\pi i(d_1-d_2)\frac{M}{k}}
\sum_{g=0}^\infty 2n^{\bm d}_g\Bigl(2\sin\frac{2\pi}{k}\Bigr)^{2g-2},
\end{align}
where we have written the arguments $(k,M)$ to express the multi-covering structure of the coefficients.
We have also introduced the extra factor of $2$ in front of the Gopakumar-Vafa invariants $n_g^{\bm d}$, as the original modified grand potentials $J_\pm(\mu)$ associated to the density matrices $[{\widehat\rho}_{\text{U}(N|N+M)}]_\pm$ are related to $\Sigma(\mu)$ and $\Delta(\mu)$ by $J_\pm(\mu)=(\Sigma(\mu)\pm\Delta(\mu))/2$.

\begin{table}[!ht]
\begin{center}
\begin{tabular}{|c||c|c|c|c|c|c|c|}
\hline
$d$&1&2&3&4&5&6&7\\
\hline\hline
$n_0^d$&$-2$&$-2$&$-6$&$-24$&$-120$&$-678$&$-4214$\\
\hline
$n_1^d$&$0$&$1$&$8$&$73$&$676$&$6279$&$58916$\\
\hline
$n_2^d$& $0$& $0$&$-2$&$-76$&$-1556$&$-26098$&$-391604$\\
\hline
$n_3^d$& $0$& $0$&  $0$&$39$&$2020$&$65984$&$1656280$\\
\hline
$n_4^d$& $0$& $0$&  $0$&$-10$&$-1586$&$-111668$&$-4916452$\\
\hline
$n_5^d$& $0$& $0$&  $0$&$1$&$756$&$132105$&$10723496$\\
\hline
$n_6^d$& $0$& $0$&  $0$&   $0$&$-212$&$-111774$&$-17629842$\\
\hline
$n_7^d$& $0$& $0$&  $0$&   $0$&$32$&$68342$&$22182896$\\
\hline
$n_8^d$& $0$& $0$&  $0$&   $0$&$-2$&$-30194$&$-21562774$\\
\hline
$n_9^d$& $0$& $0$&  $0$&   $0$&    $0$&$9530$&$16278148$\\
\hline
$n_{10}^d$& $0$& $0$&  $0$&   $0$&    $0$&$-2092$&$-9561340$\\
\hline
$n_{11}^d$& $0$& $0$&  $0$&   $0$&    $0$&$303$&$4361964$\\
\hline
$n_{12}^d$& $0$& $0$&  $0$&   $0$&    $0$&$-26$&$-1536200$\\
\hline
$n_{13}^d$& $0$& $0$&  $0$&   $0$&    $0$&$1$&$412728$\\
\hline
$n_{14}^d$& $0$& $0$&  $0$&   $0$&    $0$&    $0$&  $-82898$\\
\hline
$n_{15}^d$& $0$& $0$&  $0$&   $0$&    $0$&    $0$&  $12036$\\
\hline
$n_{16}^d$& $0$& $0$&  $0$&   $0$&    $0$&    $0$&  $-1192$\\
\hline
$n_{17}^d$& $0$& $0$&  $0$&   $0$&    $0$&    $0$&  $72$\\
\hline
$n_{18}^d$& $0$& $0$&  $0$&   $0$&    $0$&    $0$&  $-2$\\
\hline
\end{tabular}
\caption{The diagonal Gopakumar-Vafa invariants $n_g^d$ $(d=d_1+d_2)$ identified for the chirally projected theory $J_\pm(\mu_\text{eff})$.}
\label{diagGV}
\end{center}
\end{table}
For the $N_1=N_2$ case, the diagonal Gopakumar-Vafa invariants of $J_\pm(\mu_\text{eff})=(\Sigma(\mu_\text{eff})\pm\Delta(\mu_\text{eff}))/2$ is given in table \ref{diagGV}.
This matches with the results for $d=1,2,3,4$ in \cite{MS2}.
We can perform a similar analysis for the non-diagonal case $N_1\neq N_2$.
The results are summarized in tables \ref{GV12345} and \ref{GV67}.

\begin{table}[!ht]
\begin{center}
\begin{tabular}{|c||c||c||c||c|c||c|c|}
\hline
$(d_1,d_2)$&$(1,0)$&$(1,1)$&$(1,2)$&$(1,3)$&$(2,2)$&$(1,4)$&$(2,3)$\\ \hline\hline
$n^{\bm d}_0$&$-1$&$-2$&$-3$&$-4$&$-16$&$-5$&$-55$\\ \hline
$n^{\bm d}_1$&$0$&$1$&$4$&$10$&$53$&$20$&$318$\\ \hline
$n^{\bm d}_2$&$0$&$0$&$-1$&$-6$&$-64$&$-21$&$-757$\\ \hline
$n^{\bm d}_3$&$0$&$0$&$0$&$1$&$37$&$8$&$1002$\\ \hline
$n^{\bm d}_4$&$0$&$0$&$0$&$0$&$-10$&$-1$&$-792$\\ \hline
$n^{\bm d}_5$&$0$&$0$&$0$&$0$&$1$&$0$&$378$\\ \hline
$n^{\bm d}_6$&$0$&$0$&$0$&$0$&$0$&$0$&$-106$\\ \hline
$n^{\bm d}_7$&$0$&$0$&$0$&$0$&$0$&$0$&$16$\\ \hline
$n^{\bm d}_8$&$0$&$0$&$0$&$0$&$0$&$0$&$-1$\\ \hline
$n^{\bm d}_9$&$0$&$0$&$0$&$0$&$0$&$0$&$0$\\ \hline
\end{tabular}
\caption{The general Gopakumar-Vafa invariants $n_g^{\bm d}=n_g^{(d_1,d_2)}$ with $d_1+d_2=1,2,\cdots,5$ identified for the chirally projected theory $J_\pm(\mu_\text{eff})$.}
\label{GV12345}
\end{center}
\end{table}

\begin{table}[!ht]
\begin{center}
\begin{tabular}{|c||c|c|c||c|c|c|}
\hline
$(d_1,d_2)$&$(1,5)$&$(2,4)$&$(3,3)$&$(1,6)$&$(2,5)$&$(3,4)$\\ 
\hline\hline
$n^{\bm d}_0$&$-6$&$-144$&$-378$&$-7$&$-322$&$-1778$\\ \hline
$n^{\bm d}_1$&$35$&$1272$&$3665$&$56$&$3998$&$25404$\\ \hline
$n^{\bm d}_2$&$-56$&$-4860$&$-16266$&$-126$&$-22030$&$-173646$\\ \hline
$n^{\bm d}_3$&$36$&$10850$&$44212$&$120$&$72770$&$755250$\\ \hline
$n^{\bm d}_4$&$-10$&$-15476$&$-80696$&$-55$&$-158453$&$-2299718$\\ \hline
$n^{\bm d}_5$&$1$&$14654$&$102795$&$12$&$238214$&$5123522$\\ \hline
$n^{\bm d}_6$&$0$&$-9368$&$-93038$&$-1$&$-254225$&$-8560695$\\ \hline
$n^{\bm d}_7$&$0$&$4046$&$60250$&$0$&$195788$&$10895660$\\ \hline
$n^{\bm d}_8$&$0$&$-1160$&$-27874$&$0$&$-109595$&$-10671792$\\ \hline
$n^{\bm d}_9$&$0$&$211$&$9108$&$0$&$44508$&$8094568$\\ \hline
$n^{\bm d}_{10}$&$0$&$-22$&$-2048$&$0$&$-12949$&$-4767721$\\ \hline
$n^{\bm d}_{11}$&$0$&$1$&$301$&$0$&$2626$&$2178356$\\ \hline
$n^{\bm d}_{12}$&$0$&$0$&$-26$&$0$&$-352$&$-767748$\\ \hline
$n^{\bm d}_{13}$&$0$&$0$&$1$&$0$&$28$&$206336$\\ \hline
$n^{\bm d}_{14}$&$0$&$0$&$0$&$0$&$-1$&$-41448$\\ \hline
$n^{\bm d}_{15}$&$0$&$0$&$0$&$0$&$0$&$6018$\\ \hline
$n^{\bm d}_{16}$&$0$&$0$&$0$&$0$&$0$&$-596$\\ \hline
$n^{\bm d}_{17}$&$0$&$0$&$0$&$0$&$0$&$36$\\ \hline
$n^{\bm d}_{18}$&$0$&$0$&$0$&$0$&$0$&$-1$\\ \hline
$n^{\bm d}_{19}$&$0$&$0$&$0$&$0$&$0$&$0$\\ \hline
\end{tabular}
\caption{The general Gopakumar-Vafa invariants $n_g^{\bm d}=n_g^{(d_1,d_2)}$ with $d_1+d_2=6,7$ identified for the chirally projected theory $J_\pm(\mu_\text{eff})$.
}
\label{GV67}
\end{center}
\end{table}

\section{Conclusion and discussion}\label{discussion}
In this paper we have proved that the density matrix of the theories with orthosymplectic supergroups reduces to that of the theories with unitary supergroups and the chiral projections.
We have further established the identity between the total grand potential $J(\mu)$ and the sum of the projected ones $\Sigma(\mu)$.
From this identity we can derive the Gopakumar-Vafa invariants systematically without mentioning to the numerical fitting, as long as we know those for the theories with unitary supergroups.

Let us raise several questions related to our results and discuss some further directions.

Firstly, though we have found the equivalences between the chiral projections of the $\text{U}(N|N+M)$ theories and the theories with the orthosymplectic groups, the counterpart of the U$(N|N+2M)_+$ theory is still missing (see table \ref{relation}).
It is then interesting to ask the role of this sector in the context of the orientifold background.

Secondly, we shall argue possible extension of our computations.
We have introduced the states $|m\rrb$, $|m\rrp$ in section \ref{chiral} and $|m\rangle\!\rangle$ in \cite{MS2}.
It is interesting to note that these combinations appear naturally in the Weyl character formulas for Sp$(2N)$, O$(2N)$ and O$(2N+1)$ respectively (see e.g.~\cite{FH}).
Apparently, this interpretation is a good sign for generalizations of our results on partition functions into one-point functions of the BPS Wilson loops \cite{HHMO,MM}.
It is also interesting to study the non-perturbative effects of the orientifold plane for other exactly solvable Chern-Simons theories like the $(2,2)_k$ model \cite{MN1,MN3}.

Thirdly, we have found that the worldsheet instanton in the chirally projected theories fits with the Gopakumar-Vafa formula, though the interpretation of the Gopakumar-Vafa invariants in tables \ref{diagGV}, \ref{GV12345} and \ref{GV67} is unclear to us.
From our knowledge of the unprojected theories, it is natural to expect a relation to local ${\mathbb P}^1\times{\mathbb P}^1$ with a projection.
It will be interesting to understand the Gopakumar-Vafa invariants from the geometrical viewpoint.

Finally let us discuss the BPS index.
In general the worldsheet instantons are changed drastically, while the membrane instantons remain unmodified.
This indicates that, if we assume the same refined topological string expression as in the ABJM theory \cite{HMMO}, the BPS index is reshuffled with the membrane instantons kept fixed.
For example, for $(d_1,d_2)=(1,1)$, the only non-vanishing BPS index in the ABJM theory is $N^{(1,1)}_{0,\frac{3}{2}}=1$.
In the orientifold theory, if we make an ansatz that only the BPS indices for $0\le j_L\le 1/2$ and $0\le j_L+j_R\le 3/2$ are non-vanishing from the expression of the instantons, we find non-trivial relations
\begin{align}
N^{(1,1)}_{0,\frac{1}{2}}=-2+2N^{(1,1)}_{\frac{1}{2},1},\quad
N^{(1,1)}_{0,\frac{3}{2}}=1-N^{(1,1)}_{\frac{1}{2},1},\quad
N^{(1,1)}_{\frac{1}{2},0}=2-3N^{(1,1)}_{\frac{1}{2},1},
\end{align}
by matching the expressions of the instantons.
Then, apparently there are no non-negative solutions.
We would like to search for the correct analysis for the topological invariants and see the reshuffling more clearly in future.\footnote{
In studying the ABJM partition function on an ellipsoid, it was found \cite{Ha} that the BPS index is changed from that of the local ${\mathbb P}^1\times{\mathbb P}^1$ geometry to that of local ${\mathbb P}^2$ when moving the ellipsoid deformation parameter from $b=1$ to $b=\sqrt{3}$.
The change of the BPS indices may be similar to our situation.
}

\section*{Acknowledgements}
We are grateful to Masazumi Honda, Kazumi Okuyama and Takao Suyama for valuable discussions.
The work of S.M.\ is supported by JSPS Grant-in-Aid for Scientific
Research (C) \# 26400245.

\end{document}